\documentclass[conference]{IEEEtran}
\IEEEoverridecommandlockouts
\usepackage{cite}
\usepackage{amsmath,amssymb,amsfonts}
\usepackage{graphicx}
\usepackage{textcomp}
\usepackage{xcolor}
\usepackage{algorithm, algorithmicx, algpseudocode}
\usepackage{bm}
\usepackage{hero}

\DeclareMathOperator{\diag}{diag}
\DeclareMathOperator{\blkdiag}{blkdiag}
\def\argmax{{\textnormal{argmax}}}
\def\argmin{{\textnormal{argmin}}}
\def\bpsi{{\boldsymbol{\psi}}}
\def\sigs2{}
\def\bzeta{{\boldsymbol{\zeta}}}
\def\bLambda{{\boldsymbol{\Lambda}}}
\def\blambda{{\boldsymbol{\lambda}}}
\def\b0{{\mathbf{0}}}
\def\bone{{\mathbf{1}}}

\def\Re{{\textnormal{Re}}}
\def\Im{{\textnormal{Im}}}
\def\tz{{\tilde{\bz}}}

\def\tG{{\tilde{\bG}}}
\def\tE{{\tilde{\bE}}}
\def\tLam{{\tilde{\boldsymbol{\Lambda}}}}

\begin{document}

\title{Joint Precoder and Combiner Design for MMSE Distributed Beamforming with Per-Antenna Power Constraints 
\thanks{This work was supported
  by the Defense Advanced Research Projects Agency (DARPA) under contract
  D17PC00006. The views, opinions and/or findings expressed are those of the
  authors and should not be interpreted as representing the official views or
  policies of the Department of Defense or the U.S. Government.}}

\author{\IEEEauthorblockN{Riten Gupta}
\IEEEauthorblockA{\textit{UtopiaCompression Corporation} \\
Los Angeles, CA, USA \\
riten@utopiacompression.com}
\and
\IEEEauthorblockN{Han Yan, Danijela Cabric}
\IEEEauthorblockA{\textit{University of California, Los Angeles}\\
Los Angeles, CA, USA \\
yhaddint@ucla.edu, danijela@ee.ucla.edu}
%\and
%\IEEEauthorblockN{Danijela Cabric}
%\IEEEauthorblockA{\textit{University of California, Los Angeles}\\
%Los Angeles, CA, USA \\
%danijela@ee.ucla.edu}
}

\maketitle

\begin{abstract}
We consider minimum mean square error (MMSE) joint precoder and combiner design
for single and multi carrier distributed beamforming systems with nonuniform
per-antenna transmit power constraints. We show that, similar to the
maximum-gain problem, an iterative Gauss-Seidel algorithm can be used for
minimizing MSE which alternately optimizes the transmitter and receiver
coefficients.  In a single carrier system the optimum transmit coefficients are
obtained by a simple projection of the effective 
MISO channel.  In the multicarrier case with a sum-MSE objective, the
Gauss-Seidel approach is once again applicable, but the transmit coefficients
must be found by solving a quadratically constrained quadratic problem for
which we apply a dual gradient algorithm. A numerical example is presented
which shows improvement of 0.7 dB in carrier signal-to-noise ratio (SNR) 
relative to a projected eigenvector method for a
multicarrier DBF system with Rayleigh-faded multipath channels.
\end{abstract}


\section{Introduction}
\label{sec:Introduction}
Distributed beamforming (DBF), is a promising technology for increasing
communication range, providing improved signal-to-noise ratio (SNR), or
reducing probability of intercept and detection for geographically separated
groups of communicating nodes \cite{4785387}.  In a DBF system, multiple
transmitters with potentially heterogeneous hardware form an antenna array in
an ad-hoc manner.  Each transmit radio is power constrained and since the group
is heterogeneous, this imposes a set of nonuniform per-antenna power constaints
(PAPC) on the transmit array. On the receiving end, another group of radios
forms a receive array with nonuniform noise profile, again due to heterogeneous
hardware.

Several studies of multiple input multiple output (MIMO) communications with
uniform and nonuniform PAPC have appeared recently
 %\cite{8089411,8118138,7422140,7465807,6847698,6324461} 
for various scenarios and objectives. 
%Many of these studies focus solely on the precoder (transmit
%beamformer) design. 
For example, in \cite{8089411} the authors consider
precoder design for a multiple input single output (MISO) system with an outage
probability objective.  In \cite{8118138} a zero-forcing precoder is designed
for a broadcast channel with sum-rate objective.  In \cite{7422140}, transmit
beamformers are designed for a multiuser scenario with a
signal-to-leakage-plus-noise criterion.  In \cite{7465807} multiuser sum rate
is maximized, again for MISO channels.  In \cite{6847698} directional
beamforming under PAPC is considered for MISO channels.  In \cite{6324461} a
downlink cellular max-min problem is formulated for SINR over a set of users.
Joint transmit and receive beamforming optimization with PAPC has also been
studied. Optimum precoders (in terms of beamforming gain) for various combining
strategies with uniform PAPC, also called equal gain transmission (EGT) 
were presented in \cite{love2003equal}. Zheng, et. al.
\cite{zheng2007mimo} then proposed a cyclic (Gauss-Seidel) algorithm for joint
optimization of the precoder and combiner weights, again with a gain
objective. Since then, the Gauss-Seidel approach has been applied for
multiuser MIMO cellular downlinks with sum-MSE objectives in
\cite{shi2008mmse} and for multicarrier systems with arithmetic error
probability objectives in \cite{huang2011transmit}.
In this paper we consider joint precoder and combiner optimization for single
and multicarrier beamforming systems with PAPC and MSE objectives.  We note
that besides DBF systems, conventional (co-located) MIMO systems often have
PAPC too and our formulation is equally applicable in these cases.
%
%% considered. None of these studies considers wideband
%% or multicarrier systems, while most systems today are multicarrier.
%% In \cite{huang2011transmit}, multicarrier systems are considered 
%% with an approximation to the arithmetic error probability as
%% objective.
%% %function, but only suboptimal solutions are proposed. 
%% In this paper, we derive the jointly-optimal precoder and combiner 
%% for single and multiple carrier distributed MIMO systems with PAPC and 
%% MSE objectives.
%% %We formulate the joint precoder and combiner design problem with a %sum-MSE 
%% %objective and develop an iterative algorithm to solve it.

It is well known that in a narrowband MIMO beamforming system (i.e., where the
precoder and combiner matrices consist of only one column), maximum gain is
achieved when the precoder and combiner weight vectors 
%(denoted $\bz$ and $\bw$, respectively in this paper) 
lie in the directions of the channel's
dominant right and left eigenvectors, respectively. This principle is used
to guide beamforming weight selection with a total-transmit-power (across all
antennas) constraint.  However, when per-antenna constraints are imposed, the
gain-maximizing transmit and receive weights cannot be found in closed
form. Instead one can appeal to the Gauss-Seidel approach  which
involves alternately optimizing the transmitter and receiver weights until
convergence \cite{zheng2007mimo, shi2008mmse, tsai2009transmit, huang2011transmit}. With EGT constraints, and max-gain objective, the transmit weights
are found by projecting the effective MISO channel onto the set of
vectors with unit-magnitude components \cite{love2003equal}, 
while the receive weights satisfy the
maximum ratio combining (MRC) principle and lie in the direction of the
effective single input multiple output (SIMO) channel.

In this paper, we show that a Gauss-Seidel approach can be applied to find the
MMSE transmit and receive weights for both narrowband
and wideband (multicarrier) systems. We make the following contributions:
\begin{enumerate}
%% \item 
%% We extend the maximum gain under EGT analysis of \cite{love2003equal} and
%% \cite{zheng2007mimo} to the case of nonuniform PAPC.
\item 
We show that, similar to the maximum gain case, the single-carrier
MMSE transmit weights under PAPC are obtained by a simple projection
of the MISO channel
%$\bg=\bH^H \bw$, 
but only when this channel satisfies a certain norm condition.  We
then show that a Gauss-Seidel algorithm can be used to jointly
optimize the transmit and receive weights for MSE under PAPC.
\item
We find the optimal transmit weights with PAPC for a
multicarrier system with a sum-MSE objective using a dual gradient algorithm
which we then embed in a Gauss-Seidel algorithm to find the jointly optimal
transmit and receive weights with PAPC.
\end{enumerate}

We use the following notation. Boldfaced lowercase
symbols represent complex vectors and boldfaced uppercase symbols represent
complex matrices. The superscripts $^T$ and $^H$ denote transpose and 
conjugate (Hermitian)
transpose of a matrix or vector, respectively and 
$^*$ indicates the complex conjugate.
The quantity $\diag(a_1, \dots, a_n)$ is a diagonal
matrix with $a_1, \dots, a_n$ on the diagonal and
$\blkdiag(\bA_1, \dots, \bA_n)$ is a block-diagonal matrix with
$\bA_1, \dots, \bA_n$ on the diagonal. Statistical expectation is
denoted by $E[\cdot]$. The magnitude (modulus) of a complex number $a$ is
denoted $|a|$ and the angle by $\angle a$. Vector norms are represented by
 $\|\cdot \|_x$ with the subscript denoting the type of norm, and projection
operators are denoted $\bm{[} \cdot \bm{]}^x$ with the superscript indicating
the projection set. Finally, $\mathbb{R}^n_{++}$ and $\mathbb{R}^n_{+}$ 
are the positive and non-negative orthants of $\mathbb{R}^n$ 
respectively.

The remainder of the paper is organized as follows. Section
\ref{sec:system_model} presents the MIMO beamforming system model and
introduces the projection and norm operators which facilitate the analysis in
the sequel. Section \ref{sec:gain} summarizes the max-gain problem which
provides useful comparisons to the MMSE systems which we develop later.
Next in Section \ref{sec:mse} we derive the jointly optimal MMSE transmit
and receive weights for single carrier systems and Section \ref{sec:multicarrier}
extends this to multicarrier systems with sum-MSE objective. A numerical
example is presented in Section \ref{sec:example} which shows the utility
of the proposed algorithms. Finally, the paper is concluded in Section
\ref{sec:conclusion}.

\section{System Model}
\label{sec:system_model}

\subsection{PAPC MIMO Beamforming System}

The $n$-transmitter, $m$-receiver MIMO beamforming system is described by the
following equation
\begin{equation}
\hat{s} = \bw^H \bH \bz s + \bw^H \bn
\label{eq:s_hat}
\end{equation}
where $s$ and $\hat{s}$ are the complex transmitted and equalized symbols, 
respectively,
$\bn \in \mathbb{C}^m$ is the receive noise vector, $\bH$ is the $m \times n$ 
complex MIMO channel matrix, 
and $\bz \in \mathbb{C}^n$ and $\bw \in \mathbb{C}^m$
are the transmit and receive beamforming weight vectors, respectively.
The multicarrier case is modeled by 
\begin{equation*}
\hat{s}_k = \bw_k^H \bH_k \bz_k s_k + \bw_k^H \bn_k
\end{equation*}
where the subscript $k$ indicates the $k$th carrier. 
We assume a distributed 
system in which some node, which we call the {\it fusion center} in 
keeping with standard DBF terminology, 
has full channel state information (CSI). The fusion center computes the
jointly optimal transmit and receive weight vectors and feeds this information
to the other nodes. We consider per-antenna transmitter power 
constraints of the form 
\begin{equation}
|z_i| \le \sqrt{p_i}, \ i=1, \dots, n 
\label{eq:inequality_constraint}
\end{equation}
where $z_i$ is the $i$th component of the transmit weight vector and $p_i$ is
the maximum allowable power of the $i$th element. In the multicarrier case, we
constrain the total power across carriers: $\sum_k |z_{i,k}|^2 \le p_i$
where $z_{i,k}$ is the transmit weight of the $k$th carrier on antenna $i$.

\subsection{$\calP$-Projection and $\calP$-Norm}

Define the constraint vector  $\bfp = [p_1, \dots, p_n]^T$ and matrix
$\bP = \diag(p_1, \dots, p_n)$ and the total power  $p_T = \sum_i p_i$.
Define $\calP$ as the set of feasible transmit vectors which
meet the constraints (\ref{eq:inequality_constraint}) with equality 
(i.e., the boundary of the feasible set):
\begin{equation*}
\calP = \{\bz \in \mathbb{C}^n : |z_i| = \sqrt{p_i}, \ i=1, \dots, n\}.
\end{equation*}
It can be easily shown that the closest $\bz \in \calP$ to an arbitrary nonzero 
$\bx = [x_1, \dots, x_n]^T \in \mathbb{C}^n$ is
\begin{equation}
\bm{[}\bx\bm{]}^{\calP}  = \underset{\bz \in \calP}{\argmin} \| \bx - \bz \|
%%\nonumber\\
= [\sqrt{p_1} e^{j \angle x_1}, \dots, \sqrt{p_n} e^{j \angle x_n}]^T
\label{eq:projection}
\end{equation}
and thus $\bm{[}\bx\bm{]}^{\calP}$ is a projection operator onto the set $\calP$
\cite{boyd2004convex}, which we call the $\calP$-projection.
Equation (\ref{eq:projection}) is valid for % the $l_1$ or $l_2$ norm
any $\ell_p$ norm as well as the following ``weighted'' $\ell_1$ 
norm \cite{candes2008enhancing}, which we call the $\calP$-norm,
and which will facilitate further analysis.
\begin{equation*}
\| \bx \|_{\calP} = \|\bP^{1/2} \bx \|_1 = \sum_{i=1}^n \sqrt{p_i} |x_i|.
\label{eq:calpnorm}  
\end{equation*}

We note a few important facts regarding the $\calP$-projection and the
$\calP$-norm.  First, when $\bx \in \calP$, its norm is 
$\| \bx \|_{\calP} = p_T$.  When the power constraints are all unity, i.e.,
$\bP = \bI$, the $\calP$-norm is equivalent to the $\ell_1$ norm.
%$\| \cdot \|_1$. 
From (\ref{eq:projection}), the $\calP$-projection
is independent of the magnitudes of the components
of $\bx$ and thus $\bm{[}\bx\bm{]}^{\calP} = \bm{[}\alpha \bx\bm{]}^{\calP}$ for
any nonzero real $\alpha$. As with any norm,
we have $\| \alpha \bx \|_{\calP} = |\alpha| \cdot \| \bx \|_{\calP}$ for any
real $\alpha$. Finally, for any nonzero $\bx$, we have 
$\bx^H \bm{[}\bx\bm{]}^{\calP} = \| \bx \|_{\calP}$

\subsection{Constrained MMSE Problems}

In this paper, our goal is to minimize MSE subject to PAPC. From the narrowband
model (\ref{eq:s_hat}). The normalized MSE is given by
\begin{IEEEeqnarray}{rCl}
\xi(\bz, \bw; \bH, \bR_n) 
&=& 
\frac{1}{\sigma_s^2} E[|\hat{s} - s|^2] \nonumber\\
&=& 
|\bw^H \bH \bz - 1|^2  + \frac{1}{\sigma_s^2} \bw^H \bR_n \bw
\label{eq:mse}
\end{IEEEeqnarray}
where $\sigma_s^2 = E[|s|^2]$ and $\bR_n = E[\bn \bn^H]$ is the noise covariance.
When $\sigma_s^2 = 1$, the transmit power of the $i$th antenna is $|z_i|^2$
and we will make this assumption henceforth. The single-carrier constrained
MMSE problem is 
\begin{IEEEeqnarray}{cCl}
\underset{\bz \in \mathbb{C}^n, \bw \in \mathbb{C}^m}{\text{minimize}} 
&\quad& 
\xi(\bz, \bw; \bH, \bR_n)  \nonumber\\
\text{subject to} 
&\quad& 
|z_i| \le \sqrt{p_i}, \ i=1,\dots, n.
\label{eq:single_carrier_problem}
\end{IEEEeqnarray}
With multiple carriers, we minimize the sum-MSE across carriers and constrain
the total power at each antenna. Sum-MSE has been used as an objective for
multi-user systems \cite{shi2008mmse}, spatial multiplexing systems
\cite{paulraj1997space}, and multicarrier systems \cite{palomar2003convex}.
The joint optimization problem is
\begin{IEEEeqnarray}{cCl}
\underset{
  \begin{subarray}{c}
    \bz_1, \dots, \bz_K \in \mathbb{C}^n \\
    \bw_1, \dots, \bw_K \in \mathbb{C}^m    
  \end{subarray}}
%\underset{\{\bz_k\} \in \mathbb{C}^n, \{\bw_k\} \in \mathbb{C}^m}
{\text{minimize}} 
&\quad& 
\sum_{k=1}^K \xi(\bz_k, \bw_k; \bH_k, \bR_{n,k})  \nonumber\\
\text{subject to} 
&\quad& 
\sum_k |z_{i,k}|^2 \le p_i, \ i=1,\dots, n.
\label{eq:multicarrier_problem}
\end{IEEEeqnarray}
% 

%%%%%%%%%%%%%%%%%%%%%%%%%%%%%%%%%%%%%%%%%%%%%%%%%%%%%%%%%%%%%%
%  
%               Section: Max Gain
%
%%%%%%%%%%%%%%%%%%%%%%%%%%%%%%%%%%%%%%%%%%%%%%%%%%%%%%%%%%%%%
\section{Max Gain Beamforming with PAPC}
\label{sec:gain}

The problem of maximizing the gain $G = | \bw^H \bH \bz|^2$ under EGT
constraints has been studied extensively, e.g., \cite{love2003equal},
\cite{zheng2007mimo} and the focus of the present paper is MSE minimization.
However, a brief summary of the results on gain maximization is in order here
as there are many parallels to the MMSE case.  
Thus, in this section we provide this
summary and in so doing, we also extend the EGT results to the case of
nonuniform PAPC.

\subsection{Max-Gain Precoder and Combiner}

The optimum unit-norm receive weight $\bw$ with transmit weight $\bz$ is
the MRC vector
\begin{equation}
\bw = \frac{\bH \bz}{\| \bH \bz \|_2} e^{j \phi}  
\label{eq:w_unconstrained_gain}
\end{equation}
where $\phi$ is an arbitrary phase shift. With no individual transmit power 
constraints, the optimum (maximal ratio transmission)
unit-norm transmit weight vector for receive weight $\bw$ is
\begin{equation}
\bz = \frac{\bH^H \bw}{\| \bH^H \bw \|_2} e^{j \theta}  
\label{eq:z_unconstrained_gain}
\end{equation}
where $\theta$ is again arbitrary. The jointly optimal
weights are $\bw = \bu_1 e^{j\phi}, \quad \bz = \bv_1 e^{j\theta}$
%% %
%% \begin{equation}
%% \bw = \bu_1 e^{j\phi}, \quad \bz = \bv_1 e^{j\theta}
%% \end{equation}
%% %
where $\bu_1$ and $\bv_1$ are the dominant left and right eigenvectors
of $\bH$.

\subsection{Max Gain Precoder and Combiner with PAPC}

With PAPC on the transmitter, the optimum transmit vector $\bz$ with
receive vector $\bw$ is
\begin{equation}
\bz = \bm{[}\bH^H \bw\bm{]}^{\calP} e^{j\theta}
\label{eq:z_egt_gain}
\end{equation}
with arbitrary $\theta$. Note that the optimum EGT weight vector given in
\cite{love2003equal} is a special case of the above equation, with
$\bfp = \bone$.
Assuming MRC receive weights, the optimum 
transmit phase vector 
$\bpsi = [\angle z_1, \dots, \angle z_n]^T$ solves
%Joint optimization of $\bw$ and $\bz$ with equality
%transmit power constraints and MRC consists of solving the following hard problem 
%for the transmit weight vector's phases: 
$\bpsi^\star = \argmax_\bpsi \| \bH \bP^{1/2} e^{j \bpsi} \|_2$ 
(also a generalization of
a result from \cite{love2003equal}).

\subsection{Joint Gain Maximization with Gauss-Seidel Algorithm}
\label{sec:cyclic_gain}

While a closed-form solution for the jointly optimal weights has not been
found, \cite{zheng2007mimo} proposed a Gauss-Seidel (or ``cyclic'') algorithm
wherein the transmit and receive weights are alternately updated according to
(\ref{eq:w_unconstrained_gain}) and (\ref{eq:z_egt_gain}). Gauss-Seidel
algorithms are also referred to as block coordinate descent algorithms and
their convergence is studied in \cite{bertsekas1999nonlinear} and
\cite{stoica2004cyclic}. They are appealing since at every iteration, the 
objective is guaranteed to improve or stay the same.
However, convergence can only be to a local minimum
(for objectives which are non-convex in the joint set of decision variables),
and even this convergence requires somewhat restrictive conditions
\cite{bertsekas1999nonlinear, shi2008mmse}. 
Nevertheless, the Gauss-Seidel approach has been
applied successfully in several MIMO precoder/combiner optimization problems
\cite{zheng2007mimo, shi2008mmse, huang2011transmit}.  In the max-gain case,
the algorithm starts with initializing $\bw^{(0)}$, for example to
$\bu_1$. Then equations (\ref{eq:z_egt_gain}) and
(\ref{eq:w_unconstrained_gain}) are alternately applied until convergence.

%%%%%%%%%%%%%%%%%%%%%%%%%%%%%%%%%%%%%%%%%%%%%%%%%%%%%%%%%%%%%%
%  
%               Section: MSE
%
%%%%%%%%%%%%%%%%%%%%%%%%%%%%%%%%%%%%%%%%%%%%%%%%%%%%%%%%%%%%%
\section{Single-Carrier MMSE Beamforming with PAPC}
\label{sec:mse}

We now turn to the MMSE problem, which is the focus of this paper.
In this section we study the single-carrier case in depth and
in the next section we study the multicarrier case.
Our goal is to minimize the MSE subject to PAPC, i.e., to solve
problem (\ref{eq:single_carrier_problem}).
Note that the MSE (\ref{eq:mse}) is convex in $\bz$ for a fixed
$\bw$ and vice versa, but it is not convex in $(\bz,\bw)$
\cite{palomar2003convex}, which provides some motivation for
a Gauss-Seidel approach. We first summarize the MMSE problem
without PAPC.

% ***** NOTE *******************************************************
% In the remainder of the paper, we assume \sigma_s^2 = 1.
% I am replacing \sigma_s^2 with \sigs2 (which will print nothing
% but will serve as a reminder that \sigma_s^2 would belong there
% if it was not = 1).
% ******************************************************************

\subsection{MMSE Precoder and Combiner}
\label{sec:mse_uncons}

The MMSE receive weight $\bw$ for a fixed transmit weight vector $\bz$ is
\begin{equation}
\bw = 
\frac{\sigs2 \bR_n^{-1} \bH \bz}{1 + \sigs2 \bz^H \bH^H 
\bR_n^{-1} \bH \bz}.
\label{eq:mmse_optw}
\end{equation}
The optimal transmit weight vector for fixed $\bw$ is
%$\bz = \bH^H \bw \| \bH^H \bw \|_2^{-2}$
%
\begin{equation*}
\bz = \frac{\bH^H \bw}{\| \bH^H \bw \|_2^2}
\end{equation*}
which is the same as (\ref{eq:z_unconstrained_gain}) up to a constant.  
Joint optimization of $\bw$ and $\bz$ is accomplished by expressing the
resultant MSE with the optimal receive vector 
(\ref{eq:mmse_optw}) as
\begin{equation}
\xi(\bz, \bH, \bR_n) = \frac{1}
{1 + \sigs2 \bz^H \bH^H \bR_n^{-1} \bH \bz}
\label{eq:resultant_mse}
\end{equation}
which is minimized for a maximum total transmit power $p_T$
by $\bz^\star = \sqrt{p_T} \bzeta$ where $\bzeta$ is
the (unit norm) dominant eigenvector of $\bH^H \bR_n^{-1} \bH$
\cite{palomar2003convex}.

\subsection{MMSE Precoder and Combiner with PAPC}

\subsubsection{Constrained Problem}

To find the jointly optimal transmit and receive weights for the MMSE
problem with PAPC (\ref{eq:single_carrier_problem}) 
we could try to maximize the denominator of the RHS
of (\ref{eq:resultant_mse}) under per-antenna power constraints, but
the problem would not be convex.  Instead we attempt to find a formula
analogous to (\ref{eq:z_egt_gain}) which optimizes the transmitter
weights $\bz$ for fixed receive weights $\bw$ with MSE objective.
%for the optimum transmit weights $\bz$ with power
%constraints and MSE objective. 
This, along with (\ref{eq:mmse_optw}) can then be used in a cyclic
algorithm to jointly optimize $\bw$ and $\bz$ for MSE.  
%We focus on a
%problem with inequality constraints, but find that in all cases of
%interest, the solution has active constraints.  
From (\ref{eq:mse}), we see that minimizing 
$\xi(\bz, \bw; \bH, \bR_n)$ with respect to $\bz$
is equivalent to minimizing $|\bz^H \bg|^2 - 2\Re(\bz^H\bg)$, where
\begin{equation}
\bg = \bH^H \bw
\label{eq:miso}
\end{equation}
is the effective MISO channel. The transmitter optimization problem is
\begin{IEEEeqnarray}{cCl}
\underset{\bz \in \mathbb{C}^n}{\text{minimize}} 
&\quad& 
\bz^H \bG \bz - 2\Re(\bz^H\bg)  \nonumber\\
\text{subject to} 
&\quad& 
\bz^H \bE_i \bz \le p_i, \ i=1,\dots, n
\label{eq:p_complex}
\end{IEEEeqnarray}
where $\bG = \bg \bg^H$, $\bE_i = \bfe_i \bfe_i^T$ and 
$\bfe_i$ is the $i$th standard basis vector of $\mathbb{R}^n$.
In Appendix \ref{sec:app1} we show that this quadratically constrained
quadratic problem (QCQP) exhibits strong duality.
The solutions are given in the following three
propositions, which are proved in Appendix \ref{sec:app1}. These solutions
depend on the MISO channel $\bfg$ given in (\ref{eq:miso}).
\begin{propositions}
If $\| \bg \|_{\calP} \le 1$, then $\bz^\star = \bm{[}\bg\bm{]}^{\calP}$ 
is an optimal power-constrained transmit beamforming
weight vector, i.e., a solution to
(\ref{eq:p_complex}), with non-negative Lagrange multipliers
$\lambda_i^\star = |g_i| p_i^{-1/2} (1 - \| \bfg \|_{\calP})$, $i=1,\dots, n$.
\label{prop:active}
\end{propositions}

\begin{propositions}
If $\min_i |g_i| \ge (\sum_k \sqrt{p_k})^{-1}$, 
then $\bz^\star$ is a solution to (\ref{eq:p_complex})
with Lagrange multipliers $\blambda^\star = \b0$, where
$z_i^\star = \sqrt{p_i} (\sum_k \sqrt{p_k})^{-1} (g_i^*)^{-1}$.
\label{prop:inactive}
\end{propositions}

\begin{propositions}
If $|g_i| \ge n^{-1} {p_i}^{-1/2}$ for all $i=1 \dots n$,
then $\bz^\star$ is a solution to (\ref{eq:p_complex})
with Lagrange multipliers $\blambda^\star = \b0$, where
$z_i^\star = (n g_i^*)^{-1}$.
\label{prop:inactive2}
\end{propositions}

%% Problem (\ref{eq:p_complex}) can be converted to a real problem (see 
%% \cite{luo2010semidefinite}), which is a quadratically constrained
%% quadratic problem (QCQP). It is convex, and a strictly feasible $\bz$
%% exists (for example, $[\epsilon, 0, \dots, 0]^T$ with
%% $\epsilon < \sqrt{p_1}$). Thus Slater's condition is satisfied and
%% the problem admits strong duality. Therefore, the KKT conditions are 
%% sufficient for primal-dual optimality \cite{boyd2004convex} and the
%% proofs of all three propositions follow directly from the KKT conditions.

Of the solutions above, only the first (Proposition \ref{prop:active}) has
active constraints. This solution is also nearly identical to the max-gain
under PAPC solution (\ref{eq:z_egt_gain}).  However this solution is only valid
when the $\calP$-norm of the effective MISO channel $\bg=\bH^H \bw$ is less
than unity. We show next, however, that in the context of a Gauss-Seidel 
algorithm for joint optimization of $\bz$ and $\bw$, this condition is
easily managed and the solution of Proposition \ref{prop:active} can 
be used throughout the algorithm's iterations.

\begin{propositions}
If $\| \bH^H \bw \|_{\calP} > 1$, then $\bz^\star = \bm{[}\bH^H \bw\bm{]}^{\calP}$
is the optimal power-constrained transmit beamforming vector for
receive vector 
\begin{equation*}
\hat{\bw} = \frac{\bw}{\| \bH^H \bw \|_{\calP}}
\label{eq:what}
\end{equation*}
and the MSE with $\bz^\star$ and $\hat{\bw}$ is lower than any 
achievable MSE with receive vector $\bw$. That is
$\xi(\bz^\star, \hat{\bw}; \bH, \bR_n) < \xi(\bz, \bw; \bH, \bR_n), \ \forall \bz \in \mathbb{C}^n$.
\label{prop:scaled}
\end{propositions}
{\it Proof.}
Assume $\| \bH^H \bw \|_{\calP} > 1$ and define $\bz^{\star}$ and $\hat{\bw}$
as above. Since $\| \bH^H \hat{\bw} \|_{\calP} = 1$, 
and $\bz^\star = \bm{[}\bH^H \bw\bm{]}^{\calP} = \bm{[}\bH^H \hat{\bw}\bm{]}^{\calP}$,
by Proposition \ref{prop:active}, $\bz^\star$ is optimal for $\hat{\bw}$. 
Recall that $\bg^H \bm{[}\bg\bm{]}^{\calP} = \| \bg \|_{\calP}$ for any $\bg$.
Thus we have
$\hat{\bw}^H \bH \bz^\star = (\bH^H \hat{\bw})^H \bm{[} \bH^H \hat{\bw} \bm{]}^\calP =
\| \bH^H \hat{\bw} \|_{\calP} = 1$.
Therefore, from (\ref{eq:mse})
\begin{IEEEeqnarray*}{rCl}
\xi(\bz, \bw; \bH, \bR_n) &=& |\bw^H \bH \bz-1|^2 + 
%\frac{\bw^H \bR_n \bw }{\sigma_s^2} 
\bw^H \bR_n \bw % over \sigs2 (see above comment) 
\nonumber\\
\xi(\bz^\star, \hat{\bw}; \bH, \bR_n) &=& 0 +
\frac{\bw^H \bR_n \bw}{\sigs2 \| \bH^H \bw \|_{\calP}^2}.
%\qquad \Box
\end{IEEEeqnarray*}
{\mbox{} \hfill $\Box$ }

\subsubsection{Joint Transmit / Receive Optimization}

From Proposition \ref{prop:scaled}, if $\| \bH^H \bw \|_{\calP} > 1$,
we know that the jointly optimal solution cannot include $\bw$.
From Proposition \ref{prop:active},
if $\| \bH^H \bw \|_{\calP} \le 1$, $\bz = \bm{[}\bH^H \bw\bm{]}^{\calP}$ is optimal
for $\bw$. Thus the optimal pair must satisfy $\| \bH^H \bw \|_{\calP} \le 1$, 
and $\bz = \bm{[}\bH^H \bw\bm{]}^{\calP}$. Furthermore, when $\bz$ is equal to 
the $\calP$-projection, we have $\bw^H \bH \bz = \| \bH^H \bw\|_{\calP}$. Thus
the joint optimization problem is 
\begin{IEEEeqnarray}{cCl}
\underset{\bw \in \mathbb{C}^m}{\text{minimize}} 
&\quad& 
(1 - \| \bH^H \bw \|_{\calP})^2 + %\frac{1}{\sigma_s^2}
\bw^H \bR_n \bw % over \sigs2
\label{eq:joint}\\
\text{subject to} 
&\quad& 
\| \bH^H \bw \|_{\calP} \le 1.
\nonumber
\end{IEEEeqnarray}
This problem is non-convex but has a simple solution in the MISO case
($m=1$),
with receiver noise variance $\sigma_n^2$, $n \times 1$ channel
$\bh$, and scalar receive weight $w \in \mathbb{C}$. 
The MISO problem is independent of $\angle w$ and convex in $|w|$ with 
solution
%$w^\star = \sigs2 \| \bh \|_{\calP} 
%(\sigma_n^2 + \sigs2 \| \bh \|_{\calP}^2)^{-1} e^{j\phi}$
%
\begin{equation*}
w^\star = \frac{\sigs2 \| \bh \|_{\calP}}
{\sigma_n^2 + \sigs2 \| \bh \|_{\calP}^2} e^{j\phi}, \
\bz^\star = \bm{[}\bh^* e^{j\phi}\bm{]}^{\calP}
\end{equation*}
with $\phi$ arbitrary.

\subsubsection{Shadow Prices}

The resultant MSE with $\bz = \bm{[}\bg\bm{]}^{\calP}$ 
is given above in (\ref{eq:joint}) and is
differentiable as a function of $p_i$ with
$\partial \xi / \partial p_i = -\lambda_i$. The ``shadow price'' 
interpretation of $\lambda_i$ is the reduction in MSE that can be 
realized per unit of power relaxation of the
$i$th power constraint (for small relaxations) \cite{boyd2004convex}.
%Thus, if $\lambda_i = |g_i| p_i^{-1/2} (1 - \| \bg \|_{\calP})$
%is large for some $i$, then significant MSE reduction can be obtained with 
%only a slight relaxation of the power requirement for element $i$. 
The largest
shadow price occurs for the element that maximizes $|g_i|^2 / p_i$. Thus,
a low-power element which experiences large gain to the beamformed receiver
would do well to increase its power beyond its constraint, if possible.

\subsection{Gauss-Seidel MMSE Algorithm}
\label{sec:single-carrier-cyclic}

Recall that the cyclic algorithm attempts to find a jointly optimal
pair $(\bz^\star, \bw^\star)$ by sequentially optimizing $\bz$ and $\bw$.
Optimization of $\bw$ can be carried out using (\ref{eq:mmse_optw}) and
$\bz$ can be optimized using Proposition \ref{prop:active}. 
However the situation may arise where  $\| \bH^H \bw \|_{\calP} > 1$.
But from Proposition \ref{prop:scaled} we know that in such a case
the MSE can be reduced by updating $\bw$ to 
$\hat{\bw} = \bw / \| \bH^H \bw \|_{\calP}$, and then
updating $\bz$ to 
$\bm{[}\bH^H \hat{\bw}\bm{]}^{\calP}$. But since
$\bm{[}\bH^H \hat{\bw}\bm{]}^{\calP} = \bm{[}\bH^H \bw\bm{]}^{\calP}$, 
there is no need
to update $\bw$ to $\hat{\bw}$. Thus the cyclic algorithm only needs 
to sequentially update $\bz$ and $\bw$ using $\bz = \bm{[}\bH^H \bw\bm{]}^{\calP}$
and (\ref{eq:mmse_optw}). The transmit vector can be initialized to
$\bz^{(0)} = \bm{[}\bzeta\bm{]}^{\calP}$ (see Section \ref{sec:mse_uncons}).
%where $\bzeta$ is the unconstrained 
%optimal transmit vector, which lies in the direction of the dominant 
%eigenvector of $\bH^H \bR_n^{-1} \bH$.

%%%%%%%%%%%%%%%%%%%%%%%%%%%%%%%%%%%%%%%%%%%%%%%%%%%%%%%%%%%%%%
%  
%               Section: Multicarrier
%
%%%%%%%%%%%%%%%%%%%%%%%%%%%%%%%%%%%%%%%%%%%%%%%%%%%%%%%%%%%%%

\section{Multicarrier PAPC Beamforming with Sum-MSE Objective}
\label{sec:multicarrier}

In the multicarrier case, the goal is to solve problem 
(\ref{eq:multicarrier_problem}). Once again, joint optimization of
transmit and receive coefficients is a non-convex problem and we
appeal to the Gauss-Seidel method. Optimization of the receive
weights $\bw_1, \dots \bw_K$ given a set of transmit weights
$\bz_1, \dots, \bz_K$ is a separable problem. Each carrier's
receive weight can be computed using (\ref{eq:mmse_optw}). Thus,
this portion of the Gauss-Seidel algorithm is easy. It remains
to find the optimum PAPC transmit weights given a set of receive weights.
%
%The solution to the single-carrier constrained transmit vector problem
%$\bz^\star = \bm{[}\bH^H \bw\bm{]}^{\calP}$ is simple,
%but when we have many carriers with constraints on per-antenna sum
%power, the problem is more difficult. 
%In this section we consider
%this case. We let $K$ be the number of carriers and consider
%a sum-MSE objective.
The multicarrier precoder optimization problem is
%Here we consider
%sum-MSE across $K$ carriers. The problem is
%
\begin{IEEEeqnarray}{cCl}
  \underset{\bz_1, \dots, \bz_K \in \mathbb{C}^n}{\text{minimize}} 
&\quad& 
  \sum_{k=1}^K \left( \bz_k^H \bG_k \bz_k - 
  2\Re(\bz_k^H\bg_k) \right) \nonumber\\
  \text{subject to} 
&\quad& 
  %\frac{1}{2} \sum_{k=1}^K \bz_k^H \bE_i \bz_k \le \frac{p_i}{2}, \ \forall i
  \sum_{k=1}^K | z_{i,k} |^2 \le p_i, \ i=1, \dots, n.
  \label{eq:p_mc_complex}
\end{IEEEeqnarray}
%
%where subscript $k$ indicates the quantity for the $k$th carrier.
This problem is a convex QCQP and admits strong duality, as in the
single carrier case, but the constraints introduce coupling in the
variables.  A primal gradient projection algorithm could be used, but
the projection step would be difficult. On the other hand, the dual
problem has feasible region $\mathbb{R}^n_{+}$, for which projection
is simple. Thus we solve the dual problem with a gradient projection
algorithm. The dual function, derived in Appendix \ref{sec:app2}, is
\begin{equation*}
d(\blambda) = 
\left\{
\begin{tabular}{ll}
$\displaystyle{-\sum\nolimits_k \frac{\bg_k^H \bLambda^{-1} \bg_k}{1 + \bg_k^H \bLambda^{-1} \bg_k} -\blambda^T \bfp}$, & $\blambda \in \mathbb{R}^n_{++}$ \\
$-K - \blambda^T \bfp$, & $\blambda \in \partial \mathbb{R}^n_{+}$ \\
\end{tabular}
\right.
\end{equation*}
where $\bLambda = \diag(\lambda_1, \dots, \lambda_n)$.
% and $\partial \mathbb{R}^n_{+}$ is the orthant boundary.
The dual problem is
\begin{equation*}
  \underset{\blambda \in \mathbb{R}^n_{+}}{\text{maximize}} \
  d(\blambda).
  \label{eq:d_mc}
\end{equation*}
%
%with dual feasible region $\mathbb{R}^n_+$.
%Since $d(\blambda)$ is unbounded
%below on the boundary of the feasible region, we choose a closed subset
%of $\mathbb{R}^n_{++}$ for gradient projection, for example
%$\calS = \{\blambda \in \mathbb{R}^n : \min_i \lambda_i \ge \epsilon > 0\}$.
The dual function $d(\blambda)$ is concave and differentiable on
$\mathbb{R}^n_{++}$ and projection onto the orthant is simple.  Primal
variables are calculated with the Lagrangian minimizers given in
(\ref{eq:lagrangian_z_posdef}) and (\ref{eq:lagrangian_z_possemidef}).

%\subsubsection{Combined Cyclic / Dual Algorithm}

We now have the necessary tools to jointly optimize $\{(\bz_k, \bw_k)\}$.  In
Algorithm 1 optimization steps for $\{\bz_k\}$ and $\{\bw_k\}$ are alternated.
First the transmit weights are initialized to the projected dominant
eigenvectors of each carrier and then assigned equal power (Step 1). Then the
cyclic iterations begin.  Optimization of the receive weights $\{\bw_k\}$ is a
separable problem, where each carrier's weight $\bw_k$ is found using $f(\bz,
\bH, \bR_n) = \bR_n^{-1} \bH \bz (1+ \bz^{H} \bH^H \bR_n^{-1} \bH \bz)^{-1}$ as
in (\ref{eq:mmse_optw}).  Next, the dual variables $\blambda$ are initialized
and updated using gradient ascent on the dual function followed by projection
on the orthant $\mathbb{R}^n_{+}$.  The step size $\alpha_j$ is found using a
line search. Finally, the updated transmit weights are found from the
Lagrangian minimization equation (\ref{eq:lagrangian_z_posdef})
as shown in Step 9, (or (\ref{eq:lagrangian_z_possemidef}) if $\blambda$
lies on the boundary of the orthant). Note $\bLambda_j = \diag(\blambda_j)$.

\begin{algorithm}[t]
\caption{Cyclic multicarrier precoder/combiner design}
\begin{algorithmic}[1]
\label{alg:cyclic}
%\Statex \textbf{Input:} 
%\Statex \textbf{Output:} 
  %\State
  %$i \gets 0$
  \State 
  $\bz_k^{(0)} \gets \frac{1}{\sqrt{K}} \bm{[}\bzeta_k\bm{]}^{\calP}, \ \forall k$
  %\While{not converged}
  %Repeat until convergence
  %\EndWhile  
  \For{$i=1:\textnormal{max\_cyclic\_iterations}$}
  %\State 
  %$i \gets i+1$
  \State
  $\bw_k^{(i)} \gets f(\bz_k^{(i-1)}, \bH_k, \bR_n), \ \forall k$
   \State
    $\bg_k^{(i)} \gets \bH_k^H \bw_k^{(i)}, \ \forall k$
  \State
  $\blambda \gets \blambda_0$
    \For{$j=1:\textnormal{max\_dual\_iterations}$}
    \State
    $\blambda_j \gets \bm{[}\blambda_{j-1} + \alpha_j \nabla 
    d(\blambda_{j-1})\bm{]}^{+}$
    \EndFor{}%{converged or max iterations reached}
    \State
    $\bz_k^{(i)} \gets \bLambda_j^{-1} \bg_k^{(i)} (1 + \bg_k^{(i)H} 
    \bLambda_j^{-1} \bg_k^{(i)})^{-1}, \ \forall k$
  \EndFor{}%{converged or max iterations reached}
\end{algorithmic}
\end{algorithm}

%%%%%%%%%%%%%%%%%%%%%%%%%%%%%%%%%%%%%%%%%%%%%%%%%%%%%%%%%%%%%%
%  
%               Section: Numerical Example
%
%%%%%%%%%%%%%%%%%%%%%%%%%%%%%%%%%%%%%%%%%%%%%%%%%%%%%%%%%%%%%
\section{Numerical Example}
\label{sec:example}

In this section we demonstrate the utility of the cyclic algorithm 
%for multicarrier MSE minimization 
through a numerical example. We consider two
distributed groups of nodes with $n=20$ transmitters and $m=10$ receivers.
The transmit power constraints (in Watts) are chosen uniformly in 
$[0.1, 1.0]$ and the receive noise variances (in dBW) are uniform
with spread 10 dB. The noise variance is frequency independent 
so $\bR_{n,k} = \bR_n$ and the
noise between receivers is uncorrelated so $\bR_n$ is diagonal. The
system uses $K=128$ carriers over $W=10$ MHz centered at $f_c=2$ GHz. 
We assume the receivers are spaced sufficiently far apart such that for each
transmitter/receiver pair, we have an independent Rayleigh channel 
with exponential intensity profile and delay spread of 4$\mu s$.
%with 80 paths
%spaced at $\tau=(2 W)^{-1} = 50$ns (delay spread 4 $\mu s$) and exponential
%intensity profile with the 80th path 20 dB below the first.

We compare the multicarrier cyclic algorithm to three suboptimal approaches 
and to the optimum total-power-constrained weights $\bz_k^{(P_T)}$
\cite[eqn 12]{palomar2003convex}.
%well as to the 
%optimum weights $\bz_k^{(P_T)}$ with MSE objective and total power (across 
%carriers) constraint \cite{palomar2003convex}. 
The latter approach provides a lower 
bound on MSE since its feasible region contains the feasible region of 
(\ref{eq:p_mc_complex}). The first suboptimal approach uses
$\bz_k = K^{-1/2} \bm{[}\bzeta_k\bm{]}^{\calP}$. That is, for each carrier
$k$ we find the optimum unconstrained (without PAPC) weight, which is the
dominant eigenvector of $\bH_k^H \bR_n^{-1} \bH_k$, 
and project this weight on the feasible
set. No attempt is made to optimally allocate power across carriers,
and instead all carriers are allocated equal power.
%carrithe optimum unconstrained
%transmit weight is projected onto the feasible set for each carrier and
%then all carriers are allocated equal power. 
Note that this choice of transmit vectors is also used to initialize
the cyclic algorithm (Step 1 of Algorithm 1).
%(This is the intitialization step in Algorithm 1). 
In the second suboptimal method, the
single-carrier cyclic algorithm from Section \ref{sec:single-carrier-cyclic}
is run independently for each carrier $k$ and once again equal power 
is assigned to each carrier. Thus this approach should yield better
performance than the first, since each carrier's transmit and receive
weights are jointly optimized. But once again the transmit power is
not allocated optimally across carriers. Thirdly we consider 
a naive approach where we find $\bz_k^{(P_T)}$ and for every antenna
in violation of its per-antenna constraint, scale the weight magnitude 
(equally across carriers) so that the constraint is met with equality.
In implementing Algorithm 1, we terminate the cyclic loop after
20 iterations and the dual gradient loop after 200 iterations.
%In the simulations of the multicarrier cyclic algorithm, we set the maximum 
%number of cyclic iterations to
%20 and the maximum number of dual gradient iterations to 200. 

\begin{figure}
\begin{center}
\includegraphics[width=0.5\textwidth]{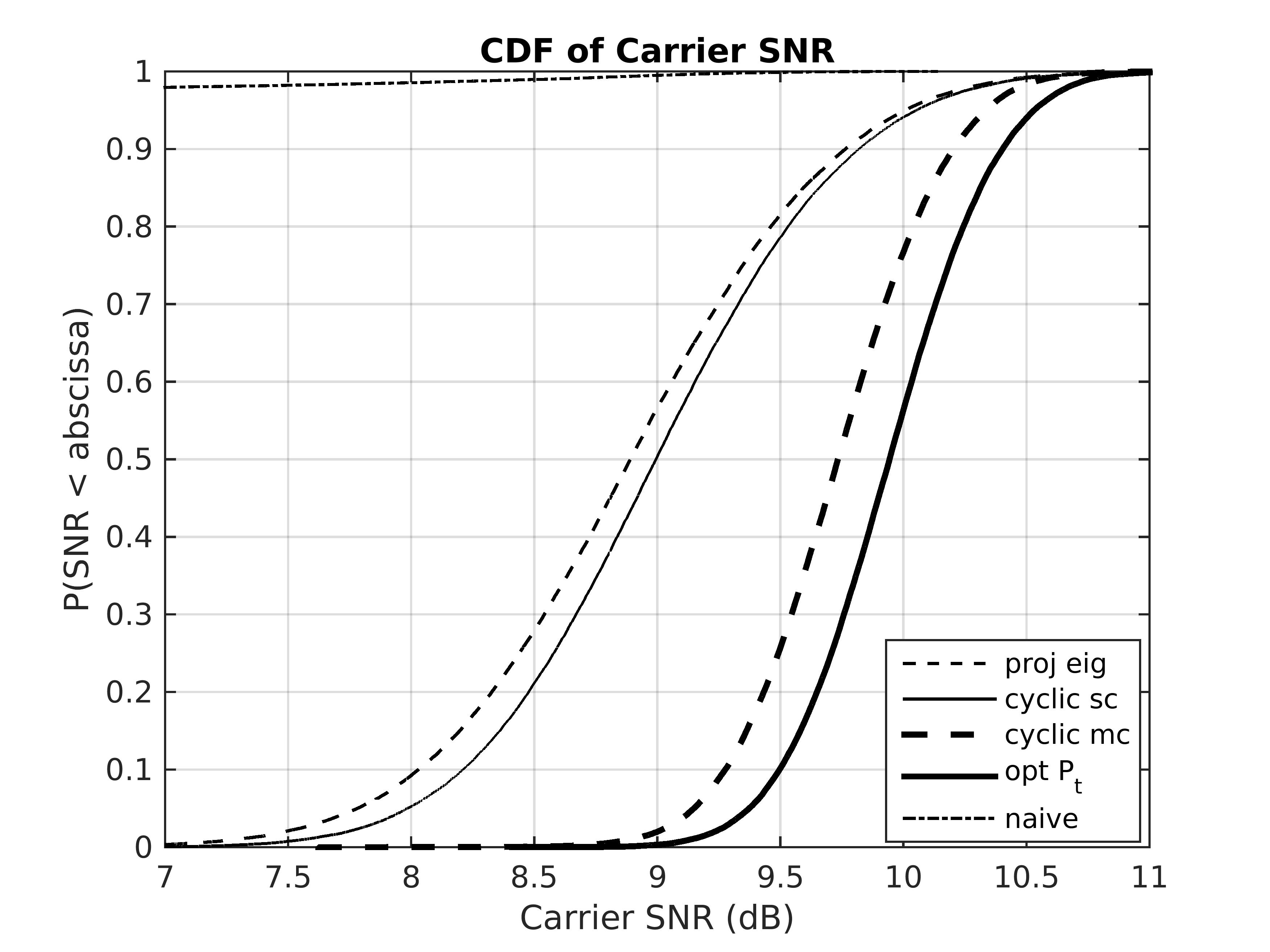}
\end{center}
\caption{Empirical CDFs of carrier SNR obtained by 
Monte-Carlo simulation with 400 trials.  }
\label{fig:snr_cdf}
\end{figure}

In Figure \ref{fig:snr_cdf} we plot the empirical cumulative
distribution functions (CDFs) of the carrier SNR, with 400 Monte-Carlo
trials, using the approaches described above. The figure shows that
the single-carrier cyclic approach is only marginally better than the
projected eigenvector approach. This is consistent with results
reported for the max-gain case which suggest that the projected
dominant eigenvector is nearly optimal for the single-carrier problem
\cite{tsai2009transmit}.  Returning to the figure, we see that the
cyclic multicarrier algorithm improves median SNR by approximately 0.7
dB relative to the first two suboptimal approaches.  The naive method
of scaling the out of tolerance weights of $\bz_k^{(P_T)}$ performs
very poorly.

\begin{figure}
\begin{center}
\includegraphics[width=0.5\textwidth]{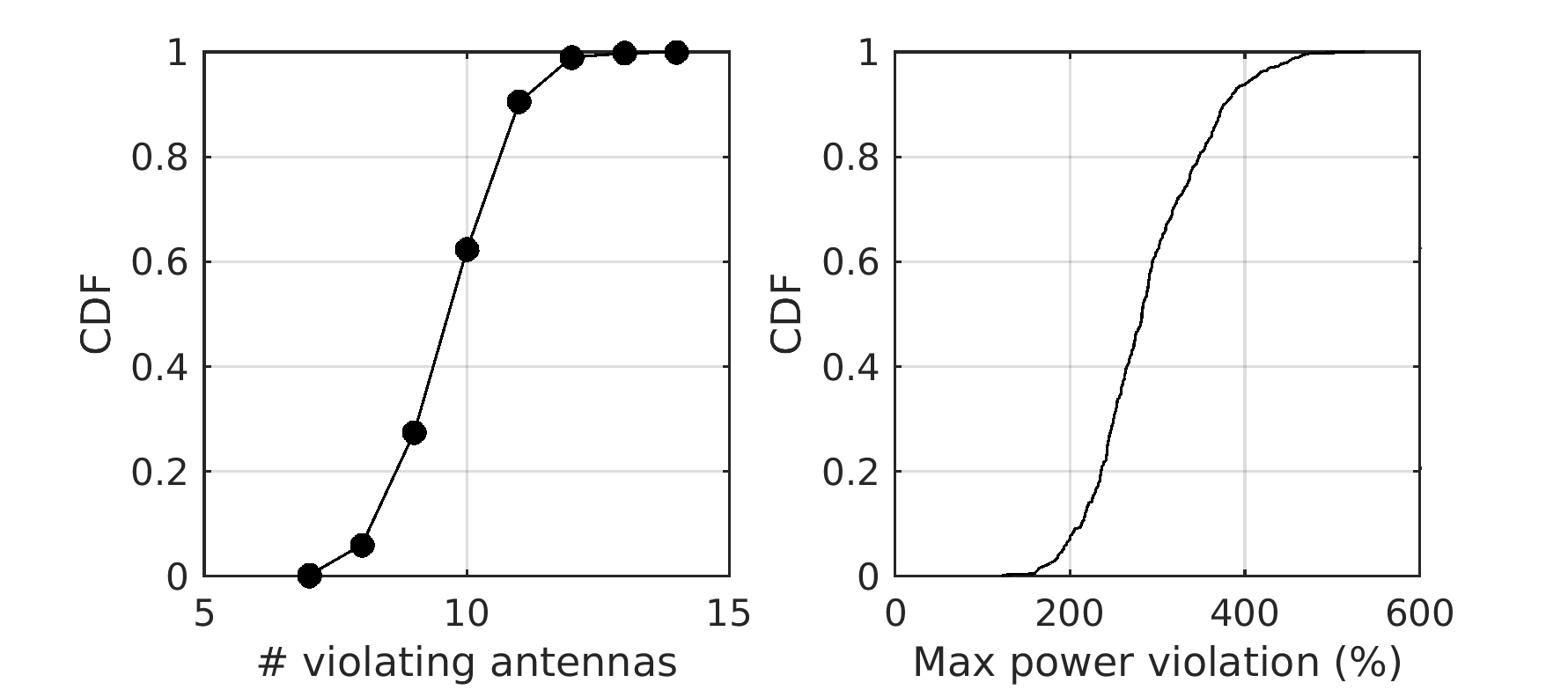}
\end{center}
\caption{Empirical CDFs of per-antenna constraint violations using 
the total-power-constrained method. Left - number of antennas in violation (out of 20).
Right - the maximum percentage violation.}
\label{fig:viol_cdfs}
\end{figure}

The total-power-constrained approach performs 0.2 dB better than the
multicarrier cyclic algorithm, but this marginal improvement comes at
a steep price in terms of per-antenna constraint violations. Figure
\ref{fig:viol_cdfs} sheds light on these violations by showing the
CDFs for the number of antennas in violation and for the maximum
percent violation.  There are always at least 7 antennas in violation
and the median value of the maximum percent violation is well over
200\%. Thus this method is not feasible with per-antenna constraints.

%yet its weights meet all power constraints, while those of $\bz_k^{(P_T)}$ do not. 
%We find that, using the latter approach,
%in all 400 Monte-Carlo trials, there are at least 7 transmitters in 
%violation of their constraints and the antenna with the largest percent violation
%always exceeds its constraint by over 123\%.

%%%%%%%%%%%%%%%%%%%%%%%%%%%%%%%%%%%%%%%%%%%%%%%%%%%%%%%%%%%%%%
%  
%               Section: conclusion
%
%%%%%%%%%%%%%%%%%%%%%%%%%%%%%%%%%%%%%%%%%%%%%%%%%%%%%%%%%%%%%
\section{Conclusion}
\label{sec:conclusion}

We have presented a method for joint transmit and receive beamforming
optimization with MSE objective and nonuniform PAPC for both single
and multicarrier systems, which have applicability for distributed
beamforming systems.  We showed that the optimum transmit weights
under PAPC are nearly identical to the gain maximizing weights under a
certain norm condition on the effective MISO channel. We also
developed a Gauss-Seidel algorithm for joint MSE optimization of the
transmitter and receiver weights, which is not encumbered by this norm
condition. Optimum transmit weights for multicarrier systems with
sum-MSE objective were found using a dual gradient algorithm
which solves a QCQP and another Gauss-Seidel algorithm was developed
for the multicarrier case. Finally, we showed through numerical
example the benefits of the method by comparison to several suboptimal
approaches. The numerical example also demonstrated the 
disadvantage of a total-power-constrained approach.

%\section*{Appendix}
\appendices
\section{Proofs of Propositions 1--3}
\label{sec:app1}

\subsection*{Proof of Proposition \ref{prop:active}}

First we note that the problem (\ref{eq:p_complex}) can be converted into 
an equivalent problem with all real variables as in \cite{luo2010semidefinite}.
Define the $\sim$ operator which maps complex vectors in $\mathbb{C}^n$
to real vectors in $\mathbb{R}^{2n}$, and complex matrices in 
$\mathbb{C}^{n \times n}$ to real matrices in $\mathbb{R}^{2n \times 2n}$
according to:
\begin{equation*}
\tilde{\bx} = \left[
  \begin{tabular}{c}
    $\Re(\bx)$ \\
    $\Im(\bx)$
  \end{tabular}
\right],
\quad
\tilde{\bA} = \left[
  \begin{tabular}{cr}
    $\Re(\bA)$ & $-\Im(\bA)$ \\
    $\Im(\bA)$ & $\Re(\bA)$ \\
  \end{tabular}
\right].
\end{equation*}
It is easy to show that $\tilde{\bA}$ is positive semidefinite if and only
if $\bA$ is positive semidefinite 
Now (\ref{eq:p_complex}) can be converted to the following real problem
\begin{IEEEeqnarray}{cCl}
\underset{\tilde{\bz} \in \mathbb{R}^{2n}}{\text{minimize}} 
&\quad& 
%\frac{1}{2} \tz^T \tilde{\bG} \tz - \tz^T \tilde{\bg} \nonumber\\
\tz^T \tilde{\bG} \tz - 2 \tz^T \tilde{\bg} \nonumber\\
\text{subject to} 
&\quad& 
%\frac{1}{2} \tz^T \tilde{\bE}_i \tz \le \frac{p_i}{2}, \ \forall i
\tz^T \tilde{\bE}_i \tz \le p_i, \ i=1, \dots, n.
\label{eq:p_real}
\end{IEEEeqnarray}
Since $\bG$ and $\bE_i$ are positive semidefinite, 
$\tilde{\bG}$ and $\tilde{\bE}_i$ are also positive semidefinite.
Problem (\ref{eq:p_real}) has a convex objective and
$n$ convex inequality constraints. Therefore it is a convex problem
\cite{boyd2004convex}. The Lagrangian function is given by
\begin{IEEEeqnarray}{rCl}
\calL(\tz, \blambda) &=& \tz^T \tG \tz -  2\tz^T \tilde{\bg}
+ \sum_{i=1}^n \lambda_i (\tz^T \tE_i \tz - p_i )
\nonumber\\
&=&
\tz^T (\tG + \tLam ) \tz
-  2\tz^T \tilde{\bfg} - \blambda^T \bfp
\label{eq:lagrangian}
\end{IEEEeqnarray}
where $\bfp = [p_1, \dots, p_n]^T$ and 
$\blambda = [\lambda_1, \dots, \lambda_n]^T$ 
are the Lagrange multipliers associated
with the magnitude constraints. Finally
$\bLambda = \diag(\lambda_1, \dots, \lambda_n)$ 
and $\tLam = \blkdiag(\bLambda, \bLambda)$.
%% %
%% \begin{equation*}
%% \tLam = \left[
%%   \begin{tabular}{cc}
%%     $\bLambda$ & $\b0$ \\
%%     $\b0$ & $\bLambda$
%%   \end{tabular}
%%   \right]
%% \end{equation*}
%% %
%% with $\bLambda = \diag(\lambda_1, \dots, \lambda_n)$. 
%
%% We have
%% %
%% \begin{equation}
%% \nabla_{\tz}\calL(\tz, \blambda) =
%% 2 (\tG + \tLam) \tz - 2 \tilde{\bg}.
%% %= 2 (\bG + \bLambda) \bz - 2 \bfg
%% \label{eq:gradient}
%% \end{equation}

Since (\ref{eq:p_real}) is a convex problem and a strictly
feasible $\tz$ exists (for example, the vector $[\epsilon, 0, \dots, 0]^T$ with
$\epsilon < \sqrt{p_1}$), Slater's condition is satisfied,
the problem admits strong duality, and 
the following Karush-Kuhn-Tucker (KKT) conditions are thus
sufficient for the primal-dual pair $(\tz^\star, \blambda^\star)$ to be 
optimal \cite{boyd2004convex}:
\begin{enumerate}
\item
$\nabla_{\tz}\calL(\tz^\star, \blambda^\star) = \b0$
\item
$\tz^\star$ satisfies all inequality constraints in (\ref{eq:p_real})
\item
$(\tilde{z}_i^{\star 2} + \tilde{z}_{i+n}^{\star 2} - p_i) \lambda_i^\star = 0, \ i=1, \dots, n$
\item
$\blambda^\star \ge \b0$
\end{enumerate}
From (\ref{eq:lagrangian}) the zero-gradient condition is 
$(\tG + \tLam) \tz = \tilde{\bg}$. Thus, in terms of complex
variables, KKT condition 1 can be written
$(\bG + \bLambda) \bz = \bfg$, or 
\begin{equation*}
(\bfg \bfg^H + \bLambda) \bz = \bfg
\end{equation*}
which is equivalent to
\begin{equation}
\lambda_i z_i = g_i ( 1 - \textstyle{\sum_k} g_k^* z_k ), \ i=1, \dots, n.
%\lambda_i z_i = g_i \left( 1 - \sum_k g_k^* z_k \right), \ i=1, \dots, n
\label{eq:zerogradient_cond}
\end{equation}
It is easily verified that this equation is satisfied for 
\begin{equation}
z_i = \sqrt{p_i} e^{j \angle g_i}, \
\lambda_i = \frac{|g_i|}{\sqrt{p_i}}(1 - \|\bfg\|_{\calP}).
\label{eq:opt_z_lambda}
\end{equation}
%
%% \begin{eqnarray}
%% z_i &=& \sqrt{p_i} e^{j \angle g_i} \nonumber\\
%% \lambda_i &=& \frac{|g_i|}{\sqrt{p_i}}(1 - \|\bfg\|_{\calP})
%% \label{eq:opt_z_lambda}
%% \end{eqnarray}
%
Furthermore, this choice of $z_i$ has magnitude $\sqrt{p_i}$. Thus the first
three KKT conditions hold. By the assumption $\|\bg\|_{\calP} \le 1$, the
Lagrange multipliers are non-negative. Thus the solution (\ref{eq:opt_z_lambda})
is optimal.
{\mbox{} \hfill $\Box$ }

\subsection*{Proof of Proposition \ref{prop:inactive}}

Letting $\lambda_i=0$ for all $i$, KKT conditions 3 and 4 are immediately
satisfied. Letting 
$z_i = \sqrt{p_i} (\sum_k \sqrt{p_k})^{-1} (g_i^\star)^{-1}$,
(\ref{eq:zerogradient_cond}) is again easily verified, satisfying
condition 1. Finally assuming
$\min_i |g_i| \ge (\sum_k \sqrt{p_k})^{-1}$, we get 
$\max_i |z_i| \le \sqrt{p_i}$, which
satisfies condition 2. This proves Proposition \ref{prop:inactive}.
{\mbox{} \hfill $\Box$ }

\subsection*{Proof of Proposition \ref{prop:inactive2}}
Apply KKT conditions again.

\section{Derivation of Multicarrier Dual Function}
\label{sec:app2}

The primal problem given in (\ref{eq:p_mc_complex}) can be converted
to a real problem using the method in Appendix \ref{sec:app1}. This problem
has Lagrangian
\begin{equation*}
\calL(\tz_1, \dots, \tz_K, \blambda) = 
\sum_{k=1}^K ( \tz_k^T (\tG_k + \tLam) \tz_k - 2 \tz_k^T \tilde{\bg}_k ) - 
\blambda^T \bfp
\end{equation*}
which is quadratic in $\tz$. The dual function is the infimum of the
Lagrangian over all transmit vectors. The matrix coefficient
$\tG_k + \tLam$ is positive definite when all Lagrange 
multipliers are positive and positive semidefinite when at least
one Lagrange multiplier is zero. (We assume dual feasibility so no
Lagrange multipliers are negative). The Lagrangian is minimized
when $(\tG_k + \tLam) \tz_k = \tilde{\bg}_k$ for all $k$, 
or, in terms of the complex variables, when 
\begin{equation}
(\bg_k \bg_k^H + \bLambda) \bz_k = \bg_k, \  k=1,\dots,K.
\label{eq:mc_zerogradient}
\end{equation}
In the positive definite case ($\blambda \in \mathbb{R}^n_{++}$) 
the minimizers are
\begin{IEEEeqnarray}{rCl}
\bz_k^\star &=& (\bg_k \bg_k^H + \bLambda)^{-1} \bg_k \nonumber\\
&=& \frac{\bLambda^{-1} \bg_k}{1 + \bg_k^H \bLambda^{-1} \bg_k}.
\label{eq:lagrangian_z_posdef}
\end{IEEEeqnarray}
Inserting these values into the Lagrangian gives the dual function
for  $\blambda \in \mathbb{R}^n_{++}$
\begin{equation*}
d(\blambda) = -\sum_k 
\frac{\bg_k^H \bLambda^{-1} \bg_k}{1 + \bg_k^H \bLambda^{-1} \bg_k}
-\blambda^T \bfp.
\end{equation*}
For the positive semidefinite case, assume there is a zero-valued 
Lagrange multiplier: $\lambda_q = 0$. Then either of the following 
sets of vectors
satisfy (\ref{eq:mc_zerogradient}) and thus minimize the Lagrangian
\begin{equation}
\bz_k^\star = \frac{1}{g_{k,q}^*} \bfe_q, \quad
\bz_k^\star = \frac{g_{k,q}}{|g_{k,q}|^2} \bfe_q, \ k=1,\dots,K
\label{eq:lagrangian_z_possemidef}
\end{equation}
(where $g_{k,q}$ is the $q$th component of $\bg_k$ and $\bfe_q$ is the $q$th
standard basis vector of $\mathbb{R}^n$). The dual function is thus
\begin{equation*}
d(\blambda) = -K - \blambda^T \bfp
\end{equation*}
when $\blambda$ is on the boundary of the non-negative orthant 
$\partial \mathbb{R}^n_+$. The dual function is
differentiable on $\mathbb{R}^n_{++}$ since the minimizers in
(\ref{eq:lagrangian_z_posdef}) are unique \cite{bazaraa2013nonlinear}.
However, we cannot claim differentiability when $\blambda$ is on the
boundary of the orthant since the minimizers in
(\ref{eq:lagrangian_z_possemidef}) are not unique.

\bibliographystyle{IEEEtran}
\bibliography{IEEEabrv,references}

\end{document}